\documentstyle[12pt]{article}
\renewcommand{\theequation}{\thesection-\arabic{equation}}
\bibliography{plain}
\title{\bf  Hierarchy of Critical Exponents on Sierpinski fractal resistor
networks} \vspace{20mm}
\author{M. A. Jafarizadeh$^{a,b,c}$ \thanks{E-mail: jafarzadeh@ark.tabrizu.ac.ir},
 M.Mirzaee$^{a,b}$\thanks{E-mail:mirzaee@ark.tabrizu.ac.ir} and
H. Aghlara$^{a,c}$\thanks{E-mail:aghlara@ark.tabrizu.ac.ir}
\\
\\
\\
$^a${\small Department of Theoretical Physics and Astrophysics, Tabriz University, Tabriz 51664, Iran.} \\
$^b${\small Institute for Studies in Theoretical Physics and Mathematics, Tehran 19395-1795, Iran.} \\
$^c${\small Pure and Applied Science Research Center, Tabriz 51664, Iran.}}

\begin{document}
\maketitle \vspace{5mm}
\begin{abstract}
Using the $S_{3}$-symmetry of Sierpinski fractal resistor networks
we determine the current distribution as well  as the
multifractals spectrum of moments of current distribution by using
the real space renormalization group technique based on
$([q/4]+1)$ independent Schure's invariant polynomials of inwards
flowing currents.\\ {\bf Keywords: Renormalization Group, Fractal,
multifractals, Resistor Network . } {\bf PACs Index: 64.60.AK and
05.50}
\end{abstract}
\vspace{70mm}
\section{INTRODUCTION}
The study of infinite sets of exponents which originated in the
field of turbulence \cite{Mandel}, has recently become the focus
of attention in a number of fields involving fractals or scaling
objects \cite{Sch,Gefen}, ranging from random resistor networks
\cite{Cle,Ste,Pak}, dynamical systems, diffusion limited
aggregates (DLA) \cite{Jul}, to localization. What is common to
all of these different fields is that, one wants to characterize
the properties of a "weight" or "measure" associated to different
parts of a fractal object. Modelization of electrical transport
properties for inhomogeneous and composite materials by random
resistor networks have been the subject of many recent works, Also
other physical phenomena such as diffusion problems can be
formulated in terms of electrical problem. Distribution of
currents (or voltage drops) on a percolating structure in the
scaling region are multifractals, in the sense that different
moments scale with different exponents, that is, if we consider a
system of length $L$, then the $q$-moment of the current
distribution :

\begin{equation}
M_q = \sum_r  I^ {\hspace{2mm} ^q}_{r}
\end{equation}

{\hspace{-5 mm}}scales as $L^{-D_{q}}$, where $D_q$  is by no
means a simple function of $q$. Thus each moment scales with its
own anomalous dimension. This phenomena is characteristic of
multifractals distributions. Actually this set of exponents first
appeared in the field of turbulence and has recently become focus
of attention in a number different fields such as diffusion
limited aggregation, dynamical system and random resistor networks
as mentioned above. Here in this paper we study the multifractals
 structure of current distribution  on Sierpinsky fractal, since as
Kirkpatric had suggested, the so called back-bone of the
percolating random resistor networks could be modeled by a fractal
structure and among the fractal objects, the $n$-simplex one is
simplest to study the various physical problems from random walk
\cite{Jaf,Jafa,Jafar,Jafar1} to electrical one on it
\cite{Ste,Pak,Jafar2,Jafar3}.

Here by using the $S_3$-symmetry of Sierpinski fractal resistor
networks (see Fig. 1) together with the minimization of the
electrical power, we have been able to determine the current
distribution in Sierpinsky fractals with decimation numbers
$b=2,3,4$, \mbox{and} $5$. Then, using the independent Shure's
$S_3$ invariant polynomials, which is proved that the required
number of independent
 Shure's $S_3$   -invariant  polynomials of degree $q$ is $[q/4]+1$,
 with [ ] indicating the greatest integer parts, we have derived the
 results of reference\cite{Ste} for $b=2$ up to $q=12$ and we have calculated $D_q$
 up to $q=22$ for $b=2,3,4 $ and \quad$5$. The organization  of the article is  as follows:

In section 2, we give a brief description of Sierpinski fractals,
then in section 3, using the $S_3$-symmetry of Sierpinsky fractal
resistor networks and minimization of electrical power we have
determined the inward flowing current of subfractals. In section 4
we talk about the independent Shure,s $S_3$-symmetry invariant
polynomials of input currents.  Section 5 is about the moments
current distributions and their multifractals spectrum where it
contains the main results of this paper. The paper ends with a
brief conclusion and 5 appendices. \vspace{3cm}
\section{ Sierpinski Fractal}
\setcounter{equation}{0} To obtain Sierpinski fractal with
decimation number $b$, we choose a triangle and divide its sides
into $b$ parts and then draw all possible lines through the
dividend points parallel to the side of the triangle. Next, having
omitted every other inner triangle, we repeat this for the
remaining triangles or for the subfractals of the next higher
order. This way Sierpinski fractals are constructed. To calculate
the fractal dimension, we label subfractals of order $(l+1)$ in
terms of partition of $(b-1)$ into $3$ positive integers
$\lambda_{_1}$, $\lambda_{_2}$ and $\lambda_{_3}$. Each partition
represents a subfractal of order $l$ and $\lambda$ shows the
distance of
 the corresponding subfractal from the  sides of triangle.
As an illustrating example, we show in Figure 2 the method of
labeling a Sierpinski fractal with decimation number $b=3$.
Obviously, the number of all possible partitions is equal to the
distribution of $(b-1)$ objects among three boxes, which is the
same as the Bose-Einstein distribution of $(b-1)$ identical bosons
in 3 quantum states. This is equal to $$
c=\frac{(b+1)!}{(b-1)!2!}. $$

According to the following definition, the fractal dimension $d_F$
of a self-similar object is $$ (N^r)^{d_F}=1, $$ with $N$ as the
number of similar objects, up to translation and rotation. For
self-similar fractals, $N$ is equal to the number of subfractals.
Therefore, we have $N=C^l$ and $r=b^{-l}$. Hence $ d_F=\frac{ln
C}{ln b} $, or $$ d_F=\frac{ln(b+1)!/(b-1)!2!}{ln b}. $$
\vspace{2cm}
\section{Determination of inward flowing current of
subfractals}
\setcounter{equation}{0}

 We denote the j-th inward flowing current of subfractal which corresponds to
 the partition $\lambda_1,\lambda_2$ and $\lambda_3$ by $I_{ \lambda_1,
 \lambda_2,\lambda_3}(\lambda_1+\delta_{1,j},\lambda_2+\delta_{2,j}
,\lambda_3+\delta_{3,j})$. Therefore, $I_1$, $I_2$ and $I_3$ can be denoted by

$$
 I_{b-1,0,0}(b,0,0),\quad I_{0,b-1,0}(0,b,0)\quad \mbox{and}\quad
 I_{0,0,b-1}(0,0,b).
$$

In order to determine the inward flowing currents in terms of
$I_{j},j=1,2$ \mbox{and} $3$,
 besides using Kirchhoff's law at each node, we have to minimize the power of
Sierpinski fractal, that is we minimize the following expression:

$$ \sum_{over\hspace{2 mm} possible\quad
partitions}\sum_{j=1}^{3}I^2_{\lambda_1,\lambda_2\lambda_3
}(\lambda_1+\delta_{1,j}, \lambda_2+\delta_{2,j},
\lambda_3+\delta_{3,j})\\$$ $$+ \mu_{\lambda_1, \lambda_2,
\lambda_3}(\sum_{j=1}^3I_{\lambda_1, \lambda_2,
\lambda_3}(\lambda_1+ \delta_{1,j}, \lambda_2+ \delta_{2,j},
\lambda_3+ \delta_{3,j})$$
\begin {equation}
+ 2\sum_{over\hspace{2 mm} all\quad nodes}\nu_{\eta_1, \eta_2,
\eta_3} (\sum_{j=1}^3I_{\eta_1, \eta_2,
\eta_3}(\eta_1-\delta_{1,j}, \eta_2-\delta_{2,j},
\eta_3-\delta_{3,j}).
\end {equation}

 {\hspace{-5 mm}}where $\mu_{\lambda_1, \lambda_2, \lambda_3}$ and $\nu_{eta_1, eta_2, \eta_3}$ are
Lagrange multipliers which are considered because of Kirchhoff,s
law on each subfractal, and also each node, respectively.
  By minimizing the energy given by expression (1-2), we get the following
  equations for the inner flowing currents:

  \begin {equation}
  I_{\lambda_1, \lambda_2, \lambda_3}
(\lambda_1+\delta_{1,j}, \lambda_2+\delta_{2,j}, \lambda_3+\delta_{3,j})
  +\mu_{\lambda_1, \lambda_2, \lambda_3}+nu_{\lambda_1+\delta_{1,j}, \lambda_2+\delta_{2,j},
  \lambda_3+\delta_{3,j}}=0,
\end {equation}

{\hspace{-5 mm}}together with the  Kirchhoff's law for each
subfractal and each vertex, respectively:

\begin {equation}
 \sum_{j=1}^{3}I_{\lambda_1, \lambda_2,
\lambda_3}(\lambda_1+\delta_{1,j},
 \lambda_2+\delta_{2,j},
\lambda_3+\delta_{3,j})=0,
\end {equation}

\begin {equation}
\sum_{j=1}^{3}I_{\eta_1, \eta_2, \eta_3}(\eta_1-\delta_{1,j},
\eta_2-\delta_{2,j}, \eta_3-\delta_{3,j})=0.
\end {equation}

Now, using  the $S_3$-permutation symmetry of Sierpinski fractal
we propose the following ansatz for the Lagrange multipliers:

\begin {equation}
\mu_{\lambda_1,\lambda_2,\lambda_{3}}=-\sum_{k=1}^{3}a_{\lambda_k}I_k,
\end{equation}
\begin{equation}
\nu_{\lambda_1,\lambda_2,\lambda_{3}}=-\sum_{k=1}^{3}b_{\lambda_k}I_k,
\end{equation}

{\hspace{-5 mm}}where $a_0$ is assumed  to be zero. Substituting
the ansatz (3-5) and (3-6) in equation (3-2), then the inward
flowing currents can be given in terms of a's and b,s ,
respectively.

\begin{equation}
 I_{\lambda_1,\lambda_2,\lambda_{3}}(\eta_1,\eta_2,\eta_{3})
=\sum_{k=1}^{3}(a_{\lambda_k}+b_{\eta_k})I_k.
\end{equation}

Actually one could write the currents in terms of input ones as in
(3-7) simply by using the  symmetry of simplex fracal  where  the
minimization of power is not required. Finally the a's and b's
themselves can be determined through the equations (3-3) and
(3-4). Here we determine the currents for b=2,3,4 and 5,
respectively. First for b=2 we have

$$
 I_{1,0,0}(2,0,0)=I_1,\quad I_{0,1,0}(0,2,0)=I_2\quad , \quad
 I_{0,0,1}(0,0,2)=I_3,
$$

$$ I_{\delta_{1,j}, \delta_{2,j},
\delta_{3,j}}(\delta_{1,j}+\delta_{1,k}, \delta_{2,j}+\delta_{2,k}
 \delta_{3,j}+\delta_{3,k})=a_1I_J+b_1I_j+b_1I_k
$$ Using equation (3-4) we obtain:

$$
   a_1+2b_1=0
$$

{\hspace{-6 mm}}and using equation (3-4) we get:

$$ 1+2a_1+b_1=0. $$\\
 Solving the above equations we get the
following result: $$ I_{\delta_{1,j}, \delta_{2,j},
\delta_{3,j}}(\delta_{1j}+\delta_{1k}, \delta_{2j}+\delta_{2k},
\delta_{3j}+\delta_{3k})=\frac{(I_k-I_j)}{3}. $$

Similarly for b=3 we have

$$ I_{2,0,0}(3,0,0)=I_1,\quad I_{0,2,0}(0,3,0)=I_2, \quad
I_{0,0,2}(0,0,3)=I_3, $$

$$ I_{2\delta_{1,j}, 2\delta_{2,j},
2\delta_{3,j}}(2\delta_{1,j}+\delta_{1,k},
2\delta_{2,j}+\delta_{2,k}, 2\delta_{3,j}+\delta_{3,k})=
a_2I_j+b_2I_j+b_1I_k, $$

$$
I_{\delta_{1,j}+\delta_{1,k},\delta_{2,j}+\delta_{2,k},\delta_{3,j}+\delta_{3,k}}
(2\delta_{1,j}+\delta_{1,k},2\delta_{2,j}+\delta_{2,k},2\delta_{3,j}+\delta_{3,k})=
a_1I_j+a_1I_k+b_2I_j+b_1I_k, $$

$$
I_{\delta_{1,j}+\delta_{1,k},\delta_{2,j}+\delta_{2,k},\delta_{3,j}+\delta_{3,k}}
(\delta_{1,j}+\delta_{1,k}+\delta_{1,l},
\delta_{2,j}+\delta_{2,k}+\delta_{2,l}
,\delta_{3,j}+\delta_{3,k}+\delta_{3,l})$$ $$
=a_1(I_j+I_k)+b_1(I_j+I_k+I_l). $$

Using equation (3-3) in subfractal
$(2\delta_{1,j},2\delta_{2,j},2\delta_{3,j})$, we get

$$ 1+2(a_2+b_2)-b_1=0. $$

Also using equation (3-3) in subfractal
$(\delta_{1,j}+\delta_{1,K},
\delta_{2,j}+\delta_{2,K},\delta_{3,j}+\delta_{3,K})$ we  get

$$ 3a_1+2b_1+b_2=0. $$

Also using equation (3-4) in the vertices we  have

$$ a_1+a_2+2b_1=0, $$

$$ a_1+2b_1=0, $$

$$ 2a_1+3b_1=0. $$

After solving the above equations we get the following result for
the currents for b=3

$$ I_{2\delta_{1,j}, 2\delta_{2,j},
2\delta_{3,j}}(2\delta_{1,j}+\delta_{1,k},
2\delta_{2,j}+\delta_{2,k}, 2\delta_{3,j}+\delta_{3,k})=
-\frac{9}{21}I_j+ \frac{3}{21}I_k, $$

$$
I_{\delta_{1,j}+\delta_{1,k},\delta_{2,j}+\delta_{2,k},\delta_{3,j}+\delta_{3,k}}
(2\delta_{1,j}+\delta_{1,k}, 2\delta_{2,j}+\delta_{2,k},
2\delta_{3,j}+\delta_{3,k})$$ $$= \frac{9}{21}I_j-\frac{3}{21}I_k,
$$

$$
I_{\delta_{1,j}+\delta_{1,k},\delta_{2,j}+\delta_{2,k},\delta_{3,j}+\delta_{3,k}}
(\delta_{1,j}+\delta_{1,k}+\delta_{1,l},\delta_{2,j}+\delta_{2,k}+\delta_{2,l},
\delta_{3,j}+\delta_{3,k}+\delta_{3,l})$$ $$
=-\frac{6}{21}(I_j+I_k)+\frac{4}{21}(I_j+I_k+I_l). $$ By the same
procedure explained  above, we can calculate the inner inward
flowing currents for decimation number $b=4$ and $b=5$, where we
quote only the results below and give the details of calculation
in Appendix I and  II. \\ \subsection{ Inner inward flowing
currents corresponding to $ b=4$:}

$$ I_{3\delta_{1,j}, 3\delta_{2,j},
3\delta_{3,j}}(3\delta_{1,j}+\delta_{1,k},
3\delta_{2,j}+\delta_{2,k}, 3\delta_{3,j}+\delta_{3,k})=
-\frac{19}{41}I_j+\frac{3}{41}I_k, $$

$$ I_{2\delta_{1,j}+\delta_{1,k}, 2\delta_{2,j}+\delta_{2,k},
2\delta_{3,j}+\delta_{3,k}}( 3\delta_{1,j}+\delta_{1,k},
\delta_{2,j}+\delta_{2,k}, 3\delta_{3,j}+\delta_{3,k})=
\frac{19}{41}I_j-\frac{3}{41}I_k, $$

$$ I_{2\delta_{1,j}+\delta_{1,k}, 2\delta_{2,j}+\delta_{2,k},
2\delta_{3,j}+\delta_{3,k}}(2\delta_{1,j}+2\delta_{1,k},
2\delta_{2,j}+2\delta_{2,k}, 2\delta_{3,j}+2\delta_{3,k})$$ $$=
-\frac{9}{41}(I_j-I_k), $$

$$ I_{2\delta_{1,j}+\delta_{1,k}, 2\delta_{2,j}+\delta_{2,k},
2\delta_{3,j}+\delta_{3,k}}
(2\delta_{1,j}+\delta_{1,k}+\delta_{1,l},
2\delta_{2,j}+\delta_{2,k}+\delta_{2,l},
2\delta_{3,j}+\delta_{3,k}+\delta_{3,l})$$ $$=
-\frac{184}{1353}I_j-\frac{52}{1353}I_k+\frac{146}{1353}I_l, $$

$$ I_{\delta_{1,j}+\delta_{1,k}+\delta_{1,l},
\delta_{2,j}+\delta_{2,k}+\delta_{2,l},
\delta_{3,j}+\delta_{3,k}+\delta_{3,l}}(2\delta_{1,j}+\delta_{1,k}+\delta_{1,l},
2\delta_{2,j}+\delta_{2,k}+\delta_{2,l},
2\delta_{3,j}+\delta_{3,k}+\delta_{3,l})$$ $$=
\frac{368}{1353}I_j-\frac{94}{1353}(I_k+I_l). $$ \vspace{1cm}
\subsection{ Inner inward flowing currents corresponding to $ b=5$:}
\setcounter{equation}{0}

$$ I_{4\delta_{1,j}, 4\delta_{2,j},
4\delta_{3,j}}(4\delta_{1,j}+\delta_{1,k},
4\delta_{2,j}+\delta_{2,k}, 4\delta_{3,j}+\delta_{3,k})=
\frac{283}{591}I_j-\frac{41375}{1015929}I_k, $$ $$
I_{3\delta_{1,j}+\delta_{1,k}, 3\delta_{2,j}+\delta_{2,k},
3\delta_{3,j}+\delta_{3,k}}( 4\delta_{1,j}+\delta_{1,k},
4\delta_{2,j}+\delta_{2,k}, 4\delta_{3,j}+\delta_{3,k})$$ $$=
-\frac{283}{591}I_j+\frac{41375}{1015929}I_k, $$ $$
I_{3\delta_{1,j}+\delta_{1,k}, 3\delta_{2,j}+\delta_{2,k},
3\delta_{3,j}+\delta_{3,k}}(3\delta_{1,j}+2\delta_{1,k},
3\delta_{2,j}+2\delta_{2,k}, 3\delta_{3,j}+2\delta_{3,k})$$ $$=
\frac{51}{197}I_j-\frac{25}{197}I_k, $$ $$
I_{3\delta_{1,j}+\delta_{1,k}, 3\delta_{2,j}+\delta_{2,k},
3\delta_{3,j}+\delta_{3,k}}
(3\delta_{1,j}+\delta_{1,k}+\delta_{1,l},
3\delta_{2,j}+\delta_{2,k}+\delta_{2,l},
3\delta_{3,j}+\delta_{3,k}+\delta_{3,l})$$ $$=
2\frac{17486}{112881}I_j+2\frac{2206}{112881}I_k
-2\frac{2448}{37627}I_l, $$

$$ I_{2\delta_{1,j}+2\delta_{1,k}, 2\delta_{2,j}+2\delta_{2,k},
2\delta_{3,j}+2\delta_{3,k}}( 3\delta_{1,j}+2\delta_{1,k},
3\delta_{2,j}+2\delta_{2,k}, 3\delta_{3,j}+2\delta_{3,k})$$ $$=
-\frac{51}{197}I_j-\frac{25}{197}I_k, $$ $$
I_{2\delta_{1,j}+2\delta_{1,k}, 2\delta_{2,j}+2\delta_{2,k},
2\delta_{3,j}+2\delta_{3,k}}(
2\delta_{1,j}+2\delta_{1,k}+\delta_{1,l},
2\delta_{2,j}+2\delta_{2,k}+\delta_{2,l},
2\delta_{3,j}+2\delta_{3,k}\delta_{3,l})$$
$$=\frac{9865}{338643}(2I_j+I_k)-\frac{12482}{338643}I_l $$ $$
I_{2\delta_{1,j}+\delta_{1,k}+\delta_{1,l},
2\delta_{2,j}+\delta_{2,k}+\delta_{2,l},
2\delta_{3,j}+\delta_{3,k}+\delta_{3,l}}
(3\delta_{1,j}+\delta_{1,k}+\delta_{1,l},
3\delta_{2,j}+\delta_{2,k}+\delta_{2,l},
3\delta_{3,j}+\delta_{3,k}+\delta_{3,l})$$ $$=
2\frac{5138}{112881}(I_k+I_l)-4\frac{34972}{112881}I_j, $$

$$ I_{2\delta_{1,j}+\delta_{1,k}+\delta_{1,l},
2\delta_{2,j}+\delta_{2,k}+\delta_{2,l},
2\delta_{3,j}+\delta_{3,k}+\delta_{3,l}}
(2\delta_{1,j}+2\delta_{1,k}+\delta_{1,l},
2\delta_{2,j}+2\delta_{2,k}+\delta_{2,l},
2\delta_{3,j}+2\delta_{3,k}+\delta_{3,l})$$ $$=
2\frac{18847}{3047787}I_j -2\frac{14356}{3047787}I_k
+\frac{624}{3047787}I_l. $$ \vspace{1cm}
\section{Shure's Polynomials of Inward Flowing Currents}
\setcounter{equation}{0}

 Shure's $S_3$-invariant polynomials are homogeneous polynomials of degree
3 of variables $I_1,I_2$ and $I_3$:

$$ s_{{\lambda_1},\lambda_2,\lambda_3}=\sum_{permutation \quad of
(1,2,\mbox{3})}
I_{1}^{\lambda_1}I_{2}^{\lambda_2}I_{3}^{\lambda_3} $$

{\hspace{-5 mm}}where $\lambda_1,\lambda_2,\lambda_3$ are
partitions of m into $3$ non-negative integers, that is:

$$
  \lambda_1+\lambda_2+\lambda_3=m.
  $$

Because of the following equation due to  Kirchhoff's law:

\begin{equation}
S_{1}=\sum_{k=1}^3I_k=0,
\end{equation}

{\hspace{-5 mm}}all  Schure's polynomials of degree m,
corresponding to all possible partitions of m, are not
independent. In calculation of the multifractals critical
exponents $D_q$, we must use the independent ones. By multiplying
both sides of (4-2) by $S_{\lambda_1,\lambda_2,\lambda_3}$, we get

\begin{equation}
 0=S_1S_{\lambda_1,\lambda_2,\lambda_3}=\sum
 a_{\mu_1,\mu_2,\mu_3}S_{\mu_1,\mu_2,\mu_3},
 \end{equation}

 {\hspace{-5mm}}where $(\mu_1,\mu_2,\mu_3)$ and $(\lambda_1,\lambda_2,\lambda_3)$ correspond to
 partition of $m$-1 and $m$ respectively. From the formula (2-10) it follows that there are $P_{3}(m+1)$ constraint
over $P_{3}(m)$ shure polynomials of degree m, where $p_{3}(m)$
takes all possible partitions of $m$ into 3 non-negative integers.
Therefore, the number of  invariant polynomials of degree m is:

\begin{equation}
P_{3}(m)-P_{3}(m-1).
\end{equation}
For example for $m=2$ we have

$$ 0=S_1S_1=S_2+2S_{1,1} $$

therefore, using the above equation we can write $S_{1,1}$ in
terms of $S_{2}$ as:
\begin{equation}
 S_{1,1}=-\frac{S_2}{2}.
\end{equation}
Thus we have only one invariant polynomial for  $q=2$. Also in the
case of $q=4$ we have

$$ S_1S_3=S_4+S_{3,1}=0,  $$ $$
S_1S_{2,1}=S_{3,1}+2S_{2,2}+S_{2,1,1}=0, $$ $$
S_1S_{1,1,1}=S_{2,1,1}=0, $$

hence there is only one independent polynomial such as $S_4$ and
the others can be written

in terms of $S_4$ as follows:

$$ S_{3,1}=-S_4,\quad\quad S_{2,2}=\frac{S_{4}}{2},$$
\begin{equation}
 S_{2,1,1}=0.
\end{equation}

For $q=6$ we have:

$$ S_1S_5=S_6+S_{5,1} =0, $$ $$
 S_1S_{4,1}=S_{5,1}+ S_{4,2}+S_{4,1,1}=0, $$ $$
S_1S_{3,2}=S_{4,2}+2S_{3,3}+S_{3,2,1}=0,$$ $$
S_1S_{3,1,1}=S_{4,1,1}+S_{3,2,1}=0.$$

Therefore, there are only two independent invariant polynomials
such as $S_{6}$, $S_{3,3}$ and the other dependent one can be
written in terms of them as follows: $$ S_{5,1} =-S_6, \hspace{5
mm}
 S_{4,2}=\frac{S_6}{2}-S_{3,3},\hspace{5 mm}
S_{321}=-\frac{S_6}{2}-S_{3,3},\hspace{5 mm}
 S_{4,1,1}=\frac{S_6}{2}+S_{3,3}\\ \\
$$

In Appendix III, we have proved that the number of independent
Schure's invariant polynomials of degree $q$ is equal to:

\begin{equation}
[q/4]+1
\end{equation}

{\hspace{-6 mm}}where [ ] means the greatest integer part.

Below we give some of the constraints over Schure's invariant
polynomials of degrees $q=8$ and $10$ which are
 occurring through imposing the Kirschhof's law
over Schure's polynomials of order eight:

$$ S_{8}+S_{7,1}=0, \quad  S_{7,1}+S_{6,2}+2S_{6,1,1}=0,\quad $$
$$S_{6,2}+S_{5,3}+S_{5,2,1}=0,\quad  S_{6,1,1}+S_{5,2,1}=0,\quad
$$ $$ S_{5,3}+2S_{4,4}+S_{4,3,1}=0, \quad
S_{5,2,1}+S_{4,3,1}+2S_{4,2,2}=0, $$

where, $S_{8}$ and $S_{4,2,2}$ are considered as the invariant
polynomials and other dependent invariant can be expressed in
terms of them as follows:
 $$ S_{7,1}=-S_8,\quad  S_{3,3,2}=S_{4,2,2},$$
$$S_{4,3,1}=-S_{4,2,2},\quad S_{5,2,1}=5S_{4,2,2}, $$ $$
S_{6,1,1}=-5S_{4,2,2},\quad  S_{6,2}=S_{8}+10S_{4,2,2}, $$ $$
S_{5,3}=-15S_{4,2,2}-S_{8},\quad
S_{4,4}=\frac{S_{8}+16S_{4,2,2}}{2}.$$

Constrains over Schure's polynomials of order ten are:

$$ S_{10}+S_{9,1}=0, \quad S_{8,2}+S_{9,1}+2S_{8,1,1}=0, $$
$$S_{8,2}+S_{7,3}+S_{7,2,1}=0, \quad S_{8,1,1}+S_{7,2,1}=0, $$
$$S_{7,3}+S_{6,4}+S_{6,3,1}=0, \quad
S_{7,2,1}+S_{6,3,1}+2S_{6,2,2},$$ $$ S_{6,4}+2S_{5,5}+S_{5,4,1}=0
\quad S_{6,3,1}+S_{5,4,1}+S_{5,3,2}=0, $$ $$ S_{6,2,2}+S_{5,3,2}=0
$$

where $S_{10}$ and $S_{4,4,2}$ are considered as the invariant
polynomials and other dependent invariants can be expressed in
terms of them as follows:
 $$ S_{4,3,3}=0,\quad S_{5,3,2}=-2S_{4,4,2},$$
$$ S_{6,2,2}=2S_{4,4,2},\quad S_{9,1}=-S_{10},$$
$$S_{5,4,1}=-S_{4,4,2},\quad S_{6,3,1}=3S_{4,4,2},$$
$$S_{7,2,1}=-7S_{4,4,2},\quad S_{8,1,1}=7S_{4.4,2},$$
$$S_{8,2}=S_{10}-14S_{4,4,2},\quad S_{7,3}=21S_{4,4,2}-S_{10},$$
$$S_{6,4}=S_{10}-24S_{4,4,2},\quad
S_{5,5}=\frac{25S_{4,4,2}-S_{10}}{2},$$

 In Appendix IV we use the
constraints concerned with the invariant polynomials of order up
to $22$ to express the dependent  invariant polynomials in terms
of the  independent ones. \vspace{3cm}
\section{Moments of Current Distribution and Multifractal Spectrum}
\setcounter{equation}{0}

In order to study the multifractals behaviour of current
distribution we  consider their $q$-moments defined as:

$$ M_q(n)=\sum_r I_r(n)^q $$

where $I_r$ is the current in the $r$-th bond of subfractals of
generation level n. From the $S_3$ symmetry of Sierpinsky fractal,
it is clear that the $q$-moments depend only on the independent
Schure's $S_{3}$ invariant polynomials of degree q of input
currents defined in section IV, that is

\begin{equation} M_q(n+1)=\sum_{ partitions\hspace{3 mm} corresponding\hspace{3
mm} to\hspace{3 mm} independent\quad
polynomials}A_{\lambda_1,\lambda_2,\lambda_3}
(n+1)S_{\lambda_1,\lambda_2,\lambda_3}(n+1), \end{equation}

{\hspace{-5 mm}}where $A_{\lambda_1,\lambda_2,\lambda_3}$,s are
some constants.

On the other hand,  $M_q(n+1)$  can be written in terms of the
invariant polynomials of their level $n$ subfractals, that is

\begin{equation}
M_q(n+1)=\sum_{partitions\quad corresponding\hspace{3 mm} to
\hspace{3 mm} invariant\quad
polynomials}A_{\lambda_1,\lambda_2,\lambda_3}(n)
S_{\lambda_1,\lambda_2,\lambda_3}(n).
\end{equation}

By comparing the expressions (5-1) and (5-2) we obtain the
recursion relations between $A_{\lambda_1,\lambda_2,\lambda_3}(n)$
and $A_{\lambda_1,\lambda_2,\lambda_3}(n+1)$. Then the scaling
factor is defined as:

\begin{equation}
\lambda(q)=\lim_{n\longrightarrow\infty}\frac{M_q(n+1)}{M_q(n)}.
\end{equation}

Obviously $\lambda(q)$ is the maximum eigenvalue of the matrix
connecting A(n) and A(n+1). Then $D(q)$, the multifractals scaling
exponents, are defined as:

\begin{equation}
D(q)=\frac{\ln(\lambda(q))}{\ln(b)},
\end{equation}

{\hspace{-5 mm}}since the $M_q(n)$ scale as:

$$ Lim_{n\longrightarrow\infty}M_q(n)=L_n^{D(q)}, $$

{\hspace{-5 mm}}where $L_n=b^n$. \\  Now, as an example, we obtain
$D_2$, the power scaling exponent of Sierpinsky fractal with
decimation numbers $b=2,3,4,4$ and $5$.

According to formula (4-4) for $q=2$ we have only one independent
invariant polynomial,
 where  we can consider $S_2$ as the independent invariant polynomial. Therefore the total
power is proportional to $S_2$, that is:

\begin{equation}
P(n+1)=A_2(n+1)S_2(n+1).
\end{equation}

It is straightforward to show that: $$
\frac{A_2(n+1)}{A_2(n)}=\frac{5}{3}  \hspace{5 mm} for\hspace{3
mm} b=2 $$ $$ \frac{A_2(n+1)}{A_2(n)}=\frac{45}{21} \hspace{5mm}
for\hspace{3 mm} b=3 $$ $$
\frac{A_2(n+1)}{A_2(n)}=\frac{3399}{1353} \hspace{5mm}
for\hspace{3 mm} b=4$$ $$
\frac{A_2(n+1)}{A_2(n)}=\frac{8576091}{3047787} \hspace{5mm}
for\hspace{3 mm} b=5 $$
 Therefore, we have:
$$ D(2)=.7369655945 \hspace{5mm} for\hspace{2mm} b=2 $$
$$D(2)=.6937297714 \hspace{5mm} for\hspace{2mm} b=3 $$
$$D(2)=.6644742613 \hspace{5mm} for\hspace{2mm} b=4 $$ $$
D(2)=.6428097998 \hspace{5mm} for\hspace{2mm} b=5$$ .

In  case of $D(4)$,we have to consider the
$S_{4}$,$S_{3,1}$,$S_{2,2}$ ,$S_{2,1,1}$, and due to relations
(4-5) only one of them, say $S_4$, is independent and the others
can be written in terms  of it. Again by computing the fourth
moments of currents of fractals which are proportional to $S_4$:

\begin{equation}
M_4(n)=A_4(n)S_4(n) \quad\quad M_4(n+1)=A_4(n+1)S_4(n+1)
\end{equation}
one can easily show that:
$$
\frac{A_{4(n+1)}}{A_{4(n)}}=1.222222222  \hspace{5mm} for\hspace{3
mm} b=2 $$
$$ \frac{A_{4(n+1)}}{A_{4(n)}}=1.288213244 \hspace{5mm}
for\hspace{3 mm} b=3 $$
 $$ \frac{A_{4(n+1)}}{A_{4(n)}}=1.323683604    \hspace{5mm}
 for\hspace{3 mm} b=4 $$
 $$ \frac{A_{4(n+1)}}{A_{4(n)}}=1.231193828     \hspace{5mm}
for\hspace{3 mm} b=5. $$

Hence using the formula (5-4) we get $$ D(4)=.2895066169
\hspace{5mm} for\hspace{2mm} b=2 $$ $$D(4)=.2305237058
\hspace{5mm} for\hspace{2mm} b=3 $$ $$D(4)=.2022791602
\hspace{5mm} for\hspace{2mm} b=4 $$ $$D(4)=.1292279056
\hspace{5mm} for\hspace{2mm} b=5 $$

{\hspace{-6 mm}}With the above explained prescription, we can
calculate the higher moments and consequently higher multifractals
exponents, where we quote only the multifractals exponents below
in the remaining part of this section and give the other
information such as the recursion relations in appendix V:
\vspace{3 cm}
\begin{table}
 \caption{The multifractals scaling exponents for $b=2,3,4,5$
 and $q=6,8,10,12,14,16,18,20,22$  }
\begin{tabular}{|c|c|c|c|c|}   \hline
 & b=2  &   b=3 & b=4 & b=5 \\ \hline
 $D(q=6)$
  &  $0.08779681671$
& $.06123675596$
  &  $.04899319186$
&$.02514677379$ \\ \hline

$D(q=8)$  &  $.02283703573$ & $.01502761610$
  &  $.01395480248$
&$.006252791011$ \\  \hline

$D(q=10)$  &  $.005703291923$ & $.003671203776$
  &  $.002879616683$
&$.001058598646$  \\  \hline

$D(q=12)$  &  $.001418367685$ & $.0009335494363$
  &  $.0007112996346$
&$.0002227018237$  \\  \hline

$D(q=14)$  &  $.0003533794924$ & $.0002242180076$
  &  $.0001768548943$
&$.00006087603594$  \\  \hline

$D(q=16)$  &  $.00008819069499$ & $.00005581142757$
  &  $.00004410760965$
&$.00001057440927$   \\ \hline

$D(q=18)$  &  $.00002202978507$ & $.00001392382295$
  &  $.00001101561388$
&$.000001496172722$   \\  \hline

$D(q=20)$  &  $.000005503871083$ & $.000003476196968$
  &  $.000002732459233$
&$.00000008760821964$   \\  \hline

$D(q=22)$  &  $.000001376330413$ & $.000002165456544$
  &  $.0000006881652063$
&$.00000005343480209$ \\  \hline
 \end{tabular}
\end{table}

  Using the above results, the
best fit we can get for various multifractals exponents are:
$$D(q,b=2)= 1 + 4\times2^{-q}, $$ $$ D(q,b=3)= 1
+51.47353178\times3^{-q}, $$ $$D(q,b=4)= 1 +
291.7913871\times4^{-q},$$ $$D(q,b=5)= 1 + 650.6706017\times
5^{-q},$$

{\hspace{-5 mm}}where the first formula is the same as the formula
of reference\cite{Ste}. The above formulas show the scaling
behaviour of the multifractals spectra. \vspace{1cm} {\large
\appendix{Appendix I: Calculation of currents of $ b=4$.}}
\\ Here in this Appendix we give the detail of the
calculation of inner inward flowing currents corresponding to
decimation number {\bf b=4}. Following the procedure of section
III, for {\bf b=4} we have: $$ I_{3\delta_{1,j}, 3\delta_{2,j},
3\delta_{3,j}}(4\delta_{1,j}, 4\delta_{2,j}, 4\delta_{3,j})=I_j,
$$
\\ $$ I_{3\delta_{1,j}, 3\delta_{2,j},
3\delta_{3,j}}(3\delta_{1,j}+\delta_{1,k},
3\delta_{2,j}+\delta_{2,k}, 3\delta_{3,j}+\delta_{3,k})=
 a_3(3)I_j+b_{31}(3)I_j+b_{31}(1)I_k,
$$  \\ $$ I_{2\delta_{1,j}+\delta_{1,k},
2\delta_{2,j}+\delta_{2,k}, 2\delta_{3,j}+\delta_{3,k}}(
3\delta_{1,j}+\delta_{1,k}, 3\delta_{2,j}+\delta_{2,k},
3\delta_{3,j}+\delta_{3,k})=
a_{21}(2)I_j+a_{21}(1)I_k+b_{31}(3)I_j+b_{31}(1)I_k, $$ \\ $$
I_{2\delta_{1,j}+\delta_{1,k}, 2\delta_{2,j}+\delta_{2,k},
2\delta_{3,j}+\delta_{3,k}}(2\delta_{1,j}+2\delta_{1,k},
2\delta_{2,j}+2\delta_{2,k}, 2\delta_{3,j}+2\delta_{3,k})=
a_{21}(2)I_j+a_{21}(1)I_k+b_{22}(2)(I_j+I_k), $$ \\ $$
I_{2\delta_{1,j}+\delta_{1,k}, 2\delta_{2,j}+\delta_{2,k},
2\delta_{3,j}+\delta_{3,k}}
(2\delta_{1,j}+\delta_{1,k}+\delta_{1,l},
2\delta_{2,j}+\delta_{2,k}+\delta_{2,l},
2\delta_{3,j}+\delta_{3,k}+\delta_{3,l})$$ $$=
a_{21}(2)I_j+a_{21}(1)+b_{211}(2)I_j+b_{211}(1)(I_k+I_l), $$  \\
$$ I_{\delta_{1,j}+\delta_{1,k}+\delta_{1,l},
\delta_{2,j}+\delta_{2,k}+\delta_{2,l},
\delta_{3,j}+\delta_{3,k}+\delta_{3,l}}(2\delta_{1,j}+\delta_{1,k}+\delta_{1,l},
2\delta_{2,j}+\delta_{2,k}+\delta_{2,l},
2\delta_{3,j}+\delta_{3,k}+\delta_{3,l})$$ $$=
{-47}a_{111}(1)(I_j+I_k+I_l)+b_{211}(2)I_j+b_{211}(1)(I_k+I_l). $$
\\ Now, imposing Kirchhoff's law on subfractals and on vertices,
we get the following equations for $\bf a$ and $\bf b$
\begin{eqnarray*}
& &1+2a_3(3)+b_{31}(3)-b_{31}(1)=0, \\ &
&3a_{21}(2)+b_{31}(3)+b_{22}(2)+2b_{211}(2)-b_{211}(1)=0, \\ &
&3a_{21}(2)+b_{31}(1)+b_{22}(2)+b_{211}(1)=0, \\ &
&3a_{111}(2)+b_{211}(2)=0\\ & &a_{21}(1)+2b_{31}(1)=0,  \\ &
&a_{3}(3)+b_{21}(2)+2b_{31}(3)=0, \\ &
&a_{21}(2)+a_{21}(1)+2b_{22}(2)=0,   \\ &
&2a_{21}(2)+a_{111}(1)+3b_{211}(2)=0, \\ &
&a_{21}(1)+a_{111}(1)+3b_{211}(1)=0,  \\ &
&3a_{111}(1)+4b_{1111}(1)=0. \\
\end {eqnarray*}
By solving the above equations we can determine inner inward
flowing currents corresponding to decimation number {\bf b=4}
which is given in subsection 3.1.
\\
\\
\\
{\large \appendix{Appendix II: Calculation of currents of {\bf
b=5}.}}
\\ Here in
 this Appendix we give the detail of the calculation of inner
 inward flowing currents corresponding to decimation number {\bf b=5}.
  Following the procedure of section III, for {\bf b=5} we have:
$$ I_{4\delta_{1,j}, 4\delta_{2,j}, 4\delta_{3,j}}(5\delta_{1,j},
5\delta_{2,j}, 5\delta_{3,j})=I_j, $$ $$ I_{4\delta_{1,j},
4\delta_{2,j}, 4\delta_{3,j}}(4\delta_{1,j}+\delta_{1,k},
4\delta_{2,j}+\delta_{2,k}, 4\delta_{3,j}+\delta_{3,k})=
 a_4(4)I_j+b_{41}(4)I_j+b_{41}(1)I_k,
$$ \\ $$ I_{3\delta_{1,j}+\delta_{1,k},
3\delta_{2,j}+\delta_{2,k}, 3\delta_{3,j}+\delta_{3,k}}(
4\delta_{1,j}+\delta_{1,k}, 4\delta_{2,j}+\delta_{2,k},
4\delta_{3,j}+\delta_{3,k})=
a_{31}(3)I_j+a_{31}(1)I_k+b_{41}(4)I_j+b_{41}(1)I_k, $$ \\ $$
I_{3\delta_{1,j}+\delta_{1,k}, 3\delta_{2,j}+\delta_{2,k},
3\delta_{3,j}+\delta_{3,k}}(3\delta_{1,j}+2\delta_{1,k},
3\delta_{2,j}+2\delta_{2,k}, 3\delta_{3,j}+2\delta_{3,k})$$ $$=
a_{31}(3)I_j+a_{31}(1)I_k+b_{32}(3)I_j+b_{32}(3)I_k,$$ $$
I_{3\delta_{1,j}+\delta_{1,k}, 3\delta_{2,j}+\delta_{2,k},
3\delta_{3,j}+\delta_{3,k}}
(3\delta_{1,j}+\delta_{1,k}+\delta_{1,l},
3\delta_{2,j}+\delta_{2,k}+\delta_{2,l},
3\delta_{3,j}+\delta_{3,k}+\delta_{3,l})$$ $$=
a_{31}(3)I_j+a_{31}(1)+b_{311}(3)I_j+b_{311}(1)(I_k+I_l), $$
\\
$$ I_{2\delta_{1,j}+2\delta_{1,k}, 2\delta_{2,j}+2\delta_{2,k},
2\delta_{3,j}+2\delta_{3,k}}( 3\delta_{1,j}+2\delta_{1,k},
3\delta_{2,j}+2\delta_{2,k}, 3\delta_{3,j}+2\delta_{3,k})=
a_{22}(2)(I_j+I_k)+b_{32}(3)I_j+b_{32}(2)I_k. $$ \\
 Again imposing
Kirchhoff's law on subfractals and on vertices, we get the
following equations for $\bf a$ and $\bf b$:
\begin {eqnarray*}
& &1+2a_4(4)+b_{41}(3)-b_{41}(1)=0, \\ &
&3a_{31}(3)+b_{41}(4)+b_{32}(3)+b_{311}(3)-b_{311}(1)=0, \\ &
&3a_{31}(1)+b_{41}(1)+b_{32}(2)=0, \\ &
&3a_{22}(1)+b_{32}(3)+b_{32}(2)+b_{221}(2)-b_{221}(1)=0, \\ &
&3a_{211}(2)+b_{311}(3)+2b_{221}(2)-b_{2111}(1)=0, \\ &
&3a_{211}(2)+b_{311}(1)+b_{221}(2)+b_{221}(1)=0, \\ &
&3a_{1111}(2)+b_{2111}(2)+3b_{2111}(1)=0, \\ &
&a_{4}(4)+a_{31}(3)+2b_{41}(4)=0,   \\ & &a_{31}(1)+2b_{41}(1)=0,
\\ & &a_{31}(3)+a_{22}(2)+2b_{32}(3)=0,  \\ &
&a_{31}(1)+a_{22}(2)+2b_{32}(2)=0, \\ &
&2a_{31}(3)+a_{211}(2)+3b_{311}(3)=0,  \\ &
&a_{31}(1)+a_{211}(1)+3b_{311}(1)=0, \\ &
&a_{22}(2)+a_{211}(2)+a_{211}(1)+3b_{221}(2)=0,   \\ &
&2a_{211}(1)+3b_{221}(1)=0, \\ &
&3a_{211}(2)+a_{1111}(1)+4b_{2111}(2)=0, \\ &
&2a_{211}(1)+a_{1111}(1)+4b_{2111}(1)=0, \\ &
&4a_{1111}(1)+5b_{11111}(1)=0. \\
\end{eqnarray*}
By solving the above equations we can determine inner inward
flowing currents corresponding to decimation number {\bf b=5}
which is given in subsection 3.2.\\

{\large \appendix{ Appendix III: Proof of the formula (4-6):}}
\renewcommand{\theequation}{\roman{section}{III}-\arabic{equation}}
\setcounter{equation}{0}
 \\ Here we give the proof of the formula (4-6). The number of independent
Shure's invariant polynomials of degree $2k$ of 3 variables
$I_1$,$I_2$,$I_3$ with the constraint:
\begin{equation}
I_1+I_2+I_3=0,
\end{equation}
is equal to
\begin{equation}
P_3(2K)-P_3(2k-1),
\end{equation}
where $P_3(m)$ is the number of partition of $m$ into at most
three independent non-negative integers. If we define $M_k(n)$,the
number of partitions of $n$ into exactly $k$ non-negative
integers, then we have
\begin {equation}
P_3(n)=\sum_{k=1}^3M_k(n).
\end{equation}
Obviously
\begin{equation}
M_1(2k)=M_1(2k-1).
\end{equation}
and
\begin{equation}
M_2(2k)=M_2(2k-1)+1 ,\quad M_2(2k)=M_2(2k+1). \\
\end{equation}
If we denote the partition of $2k$, $2k-1$ and $2k+1$ into two
non-negative integers respectively by: ($l_1,l_2$), ($m_1,m_2$)
and ($n_1,n_2$) then in the case of $l_2=m_2=n_2$ we will have
\begin{equation}
l_1=m_1-1=n_1+1.
\end{equation}
Therefore, for all values of  $l_1>k$, there is a one to one
correspondence between the $M_2(2k)$,$M_2(2k-1)$ and $M_2(2k+1)$.
Only for $l_1=k$,  $n_1$ can be equal to $k+1$, but $m_1$ cannot
be equal to $k-1$, thereof the relations (III-4) and (III-5)
follows. Now, we are ready  to prove that
\begin {equation}
M_3(2k)=M_3(2k-1)+[k/3].
\end{equation}
If we denote the partition of $2k$ and $2k-1$ into three
non-negative integers by $(l_1,l_2,l_3)$ and $(m_1,m_2,m_3)$
respectively, then using the relations (III-4) and (III-5), we can
prove
 $M_3(2k)=M_3(2k-1)$for $m_1=l_1=odd $; and for  $l_1=m_1=even$, we would have
$M_3(2k)=M_3(2k-1)$. Since $l_1$ takes values between $1$ and
$[2k/3]$, where $[[2k/3]/2]=[k/3]$ of them correspond to even
values of $l_1$, the relation (III-7) follows and the proof is
complete. \vspace{1cm} {\large \appendix{ APPENDIX IV:Solution of
Constraints Over Schure's invariants polynomials:}}

Here in this appendix by solving the constraints over Schure's
polynomials of degree $12$, $14$,$16$,$18$,$20$ and $22$, we have
expressed the dependent invariant polynomials in terms of the
independent invariant polynomials.

 1)  Solution of Constraints of degree $12$:

 The invariant polynomials  $S_{12}$, $S_{8,2,2}$ and
$S_{6,3,3}$ are considered to be independent  and the other
dependent invariant polynomials can be written in terms of them as
follows: $$ S_{5,4,3}=-S_{6,3,3},\hspace{3 mm}
S_{4,4,4}=\frac{S_{6,3,3}}{3}$$ $$ S_{7,3,2}=-S_{8,2,2},\hspace{3
mm} S_{6,4,2}=S_{8,2,2}-2S_{6,3,3}, $$ $$
S_{5,5,2}=\frac{3S_{6,3,3}-S_{8,2,2}}{2},\hspace{3 mm}
S_{6,5,1}=\frac{S_{8,2,2}-3S_{5,3,3}}{2}$$ $$
S_{7,4,1}=\frac{7S_{6,3,3}-3S_{8,2,2}}{2},\hspace{3
mm}S_{8,3,1}=\frac{5S_{8,2,2}-7S_{6,3,3}}{2}, $$ $$
S_{9,2,1}=\frac{7S_{6,3,3}-9S_{8,2,2}}{2},\hspace{3 mm}
S_{10,1,1}=\frac{9S_{8,2,2}-7S_{6,3,3}}{2}, $$ $$
S_{11,1}=-S_{12},$$ $$
 S_{10,2}=S_{12}+7S_{6,3,3}-9S_{8,2,2}, $$
$$ S_{9,3}=\frac{27S_{8,2,2}-21S_{6,3,3}-2S_{12}}{2},$$ $$
S_{8,4}=\frac{28S_{6,3,3}-32S_{8,2,2}+2S_{12}}{2},$$ $$
S_{7,5}=\frac{35S_{8,2,2}-2S_{12}-35S_{6,3,3}}{2}, $$ $$
S_{6,6}=\frac{2S_{12}+38S_{6,3,3}-36S_{8,2,2}}{4}. $$ 2) Solution
of Constraints of degree 14:

 The invariant polynomials $S_{14}$,
$S_{6,6,2}$ and $S_{5,5,4}$ are considered to be independent one
and the other dependent invariant polynomials can be written in
terms of them as follows: $$
 S_{10,3,1}=7S_{6,2,2}-23S_{5,5,4}, $$
$$ S_{8,5,1}=3S_{6,6,2}-S_{5,5,4}, $$ $$
S_{9,4,1}=-5S_{6,6,2}+7S_{5,5,4}, $$ $$ S_{8,3,3}=-5S_{5,5,4},
\hspace{3 mm}  S_{7,4,3}=5S_{5,5,4}, $$ $$ S_{6,5,3}=-S_{5,5,4},
\hspace{3 mm} S_{6,4,4}=-2S_{5,5,4},$$ $$
 S_{9,5}=-S_{14}+45S_{6,6,2}-195S_{5,5,4}, $$ $$
 S_{10,4}=S_{14}-40S_{6,6,2}+188S_{5,5,4}, $$ $$
 S_{11,3}=-S_{14}+33S_{6,6,2}-165S_{5,5,4}, $$ $$
 S_{7,7}=\frac{-S_{14}+49S_{6,6,2}-196S_{5,5,4}}{2}, $$ $$
 S_{8,6}=S_{14}-48S_{6,6,2}+196S_{5,5,4}, $$ $$
 S_{13,1}=-S_{14}, $$ $$
 S_{8,4,2}=2S_{6,62}-6S_{5,5,4}, $$ $$
 S_{12,2}=S_{14}-22S_{6,6,2}+110S_{5,5,4}, $$ $$
 S_{9,3,2}=-2S_{6,6,2}+16S_{5,5,4}, $$ $$
 S_{7,5,2}=-2S_{6,6,2}+S_{5,5,4}, $$ $$
 S_{10,2,2}=2S_{6,6,2}-16S_{5,5,4}, $$ $$
 S_{7,6,1}=-S_{6,6,2}, $$ $$
S_{11,2,1}=-11S_{6,6,2}+55S_{5,5,4}, $$ $$
 S_{12,1,1}=11S_{6,6,2}-55S_{5,5,4}. $$
3)  Solution of Constraints of degree 16: The invariant
polynomials $S_{16}$, $S_{7,7,2}$, $S_{6,6,4}$ are considered to
be independent  and the other dependent invariant polynomials can
be written in terms of them as follows: $$ S_{6,5,5}=0 $$ $$
S_{7,5,4}=-2S_{6,6,4},\hspace{3 mm}
 S_{7,6,3}=-S_{6,6,4}, $$ $$
 S_{8,8,4}=2S_{6,6,4},\hspace{3 mm}
 S_{8,5,3}=3S_{6,6,4}, $$ $$
 S_{8,7,1}=-S_{7,7,2},\hspace{3 mm}
 S_{8,6,2}=-2S_{7,7,2}+S_{6,6,4}, $$ $$
 S_{9,4,3}=-7S_{6,6,4},\hspace{3 mm}
 S_{9,5,2}=2S_{7,7,2}-4S_{6,6,4},$$ $$
 S_{9,6,1}=3S_{7,7,2}-S_{6,6,4},\hspace{3 mm}
 S_{10,3,3}=7S_{6,6,4}, $$ $$
 S_{10,4,2}=-2S_{7,7,2}+11S_{6,6,4},$$ $$
S_{10,5,1}=-5S_{7,7,2}+5S_{6,6,4},$$ $$
 S_{11,3,2}=2S_{7,7,2}-25S_{6,6,4}, $$ $$
 S_{11,4,1}=7S_{7,7,2}-16S_{6,6,4}, $$ $$
 S_{12,2,2}=-2S_{7,7,2}+25S_{6,6,4}, $$ $$
 S_{12,3,1}=-9S_{7,7,2}+41S_{6,6,4}, $$ $$
 S_{13,2,1}=13S_{7,7,2}-91S_{6,6,4}, $$ $$
 S_{14,1,1}=-13S_{7,7,2}+91S_{6,6,4}, $$ $$
 S_{15,1}=-S_{16}, $$ $$
 S_{14,2}=S_{16}+26S_{7,7,2}-182S_{6,6,4}, $$ $$
 S_{13,3}=-S_{16}-39S_{7,7,2}+273S_{6,6,4}, $$ $$
 S_{12,4}=S_{16}+48S_{7,7,2}-314S_{6,6,4},$$ $$
 S_{11,5}=-S_{16}-55S_{7,7,2}+330S_{6,6,4},$$ $$
 S_{10,6}=S_{16}+60S_{7,7,2}-335S_{6,6,4}, $$ $$
 S_{9,7}=-S_{16}-63S_{7,7,2}+336S_{6,6,4}, $$ $$
 S_{8,8}=\frac{S_{16}+64S_{7,7,2}-336S_{6,6,4}}{2}. $$

4)  Solution of Constraints of degree 18:

 The invariant
polynomials $S_{18}$, $S_{14,2,2}$, $S_{7,7,4}$, $S_{8,5,5}$ are
considered to be independent  and the other dependent invariant
polynomials can be written in terms of them as follows: $$
S_{7,6,5}=-S_{8,5,5},\hspace{3 mm}
 S_{6,6,6}=\frac{S_{8,5,5}}{3}, $$ $$
 S_{8,7,3}=-S_{7,7,4},\hspace{3 mm}
 S_{8,6,4}=S_{8,5,5}-2S_{7,7,4}, $$ $$
 S_{9,5,4}=-3S_{8,5,5}+2S_{7,7,4},\hspace{3 mm}
 S_{9,6,3}=3S_{7,7,4}-S_{8,5,5}, $$ $$
 S_{10,5,3}=4S_{8,5,5}-5S_{7,7,4},\hspace{3 mm}
 S_{10,4,4}=3S_{8,5,5}-2S_{7,7,4}, $$ $$
 S_{11,4,3}=9S_{7,7,4}-10S_{8,5,5},\hspace{3 mm}
 S_{12,3,3}=10S_{8,5,5}-9S_{7,7,4}, $$ $$
 S_{13,3,2}=-S_{14,2,2}, $$ $$
 S_{12,4,2}=S_{14,2,2}-20S_{8,5,5}+18S_{7,7,4}, $$ $$
 S_{11,5,2}=-S_{14,2,2}+30S_{8,5,5}-27S_{7,7,4}, $$ $$
 S_{10,6,2}=32S_{7,7,4}-34S_{8,5,5}+S_{14,2,2},$$ $$
 S_{9,7,2}=35S_{8,5,5}-35S_{7,7,4}-S_{14,2,2}, $$ $$
 S_{8,8,2}=\frac{36S_{7,7,4}+S_{14,2,2}-35S_{8,5,5}}{2}, $$ $$
 S_{9,8,1}=\frac{35S_{8,5,5}-S_{14,2,2}-36S_{7,7,4}}{2},$$ $$
 S_{10,7,1}=\frac{106S_{7,7,4}-105S_{8,5,5}+3S_{14,2,2}}{2}, $$
 $$
S_{11,6,1}=\frac{173S_{8,5,5}-170S_{7,7,4}-5S_{14,2,2}}{2}, $$
 $$
 S_{12,5,1}=\frac{224S_{7,7,4}-233S_{8,5,5}+7S_{14,2,2}}{2}, $$
 $$
 S_{13,4,1}=\frac{273S_{8,5,5}-260S_{7,7,4}-9S_{14,2,2}}{2}, $$
 $$
 S_{14,3,1}=260S_{7,7,4}-273S_{8,5,5}+11S_{14,2,2},$$ $$
 S_{15,2,1}=\frac{273S_{8,5,5}-260S_{7,7,4}-15S_{14,2,2}}{2}, $$
 $$
 S_{16,1,1}=\frac{260S_{7,7,4}+15S_{14,2,2}-273S_{8,5,5}}{2}, $$
 $$
 S_{17,1}=-S_{18}, $$ $$
 S_{16,2}=S_{18}-260S_{7,7,4}-15S_{14,2,2}+273S_{8,5,5}, $$ $$
 S_{15,3}=\frac{780S_{7,7,4}-2S_{18}+45S_{14,2,2}-819S_{8,5,5}}{2},
 $$ $$
 S_{14,4}=\frac{1092S_{8,5,5}-56S_{14,2,2}+2S_{18}-1040S_{7,7,4}}{2},
 $$ $$
 S_{13,5}=\frac{65S_{14,2,2}-2S_{18}-1365S_{8,5,5}+1300S_{7,7,4}}{2},
 $$ $$
 S_{12,6}=\frac{2S_{18}+1598S_{8,5,5}-1524S_{7,7,4}-72S_{14,2,2}}{2},
 $$ $$
 S_{11,7}=\frac{1694S_{7,7,4}+77S_{14,2,2}-2S_{18}-1771S_{8,5,5}}{2},
 $$ $$
 S_{10,8}=\frac{2S_{18}+1876S_{8,5,5}-80S_{14,2,2}-1800S_{7,7,4}}{2},
 $$ $$
 S_{9,9}=\frac{81S_{14,2,2}+1836S_{7,7,4}-2S_{18}-1911S_{8,5,5}}{4}.
 $$
5)  Solution of Constraints of degree 20:

 The invariant
polynomials $S_{20}$, $S_{16,2,2}$, $S_{8,8,4}$ and  $S_{7,7,6}$
are considered to be independent  and the other dependent
invariant polynomials can be written in terms of them as follows:
 $$ S_{8,8,6}=-2S_{7,7,6},\hspace{3 mm}
 S_{8,7,5}=-S_{7,7,6}, $$ $$
 S_{9,6,5}=5S_{7,7,6},\hspace{3 mm}
 S_{9,8,3}=-S_{8,8,4}, $$ $$
 S_{9,7,4}=S_{7,7,6}-2S_{8,8,4},\hspace{3 mm}
 S_{10,5,5}=-5S_{7,7,6}, $$ $$
 S_{10,6,4}=-6S_{7,7,6}+2S_{8,8,4},\hspace{3 mm}
 S_{10,7,3}=3S_{8,8,4}-S_{7,7,6}, $$ $$
 S_{11,6,3}=-5S_{8,8,4}+7S_{7,7,6},\hspace{3 mm}
  S_{11,5,4}=+16S_{7,7,6}-2S_{8,8,4}, $$ $$
 S_{12,4,4}=(-16S_{7,7,6}+2S_{8,8,4}),\hspace{3 mm}
  S_{12,5,3}=-23S_{7,7,6}+7S_{8,8,4}, $$ $$
 S_{13,4,3}=55S_{7,7,6}-11S_{8,8,4},\hspace{3 mm}
S_{14,3,3}=-55S_{7,7,6}+11S_{8,8,4}, $$ $$
 S_{15,3,2}=S_{16,2,2}, $$ $$
S_{14,4,2}=-22S_{8,8,4}+S_{16,2,2}+110S_{7,7,6},$$ $$
S_{13,5,2}=-165S_{7,7,6}-S_{16,2,2}+33S_{8,8,4},$$ $$
 S_{12,6,2}=-40S_{8,8,4}+188S_{7,7,6}+S_{16,2,2},$$ $$
 S_{11,7,2}=-195S_{7,7,6}+45S_{8,8,4}-S_{16,2,2},$$ $$
 S_{10,8,2}=-48S_{8,8,4}+196S_{7,7,6}+S_{16,2,2},$$ $$
 S_{9,9,2}=\frac{-196S_{7,7,6}+49S_{8,8,4}-S_{16,2,2}}{2},$$ $$
 S_{10,9,1}=\frac{S_{16,2,2}+196S_{7,7,6}-49S_{8,8,4}}{2},$$ $$
 S_{11,8,1}=\frac{-588S_{7,7,6}+145S_{8,8,4}-3S_{16,2,2}}{2}, $$
 $$
 S_{12,7,1}=\frac{+978S_{7,7,6}-235S_{8,8,4}+5S_{16,2,2}}{2}, $$
 $$
 S_{13,6,1}=\frac{-1354S_{7,7,6}+315S_{8,8,4}-7S_{16,2,2}}{2},$$
 $$
 S_{14,5,1}=\frac{9S_{16,2,2}-381S_{8,8,4}+1684S_{7,7,6}}{2}, $$
 $$
 S_{15,4,1}=\frac{425S_{8,8,4}-1904S_{7,7,6}-11S_{16,2,2}}{2}, $$
 $$
 S_{16,3,1}=\frac{13S_{16,2,2}+1904S_{7,7,6}-425S_{8,8,4}}{2},
 $$ $$
S_{17,2,1}=\frac{-1904S_{7,7,6}+425S_{8,8,4}-17S_{16,2,2}}{2},$$
$$
 S_{18,1,1}=\frac{17S_{16,2,2}-425S_{8,8,4}+1904S_{7,7,6}}{2}, $$
 $$
 S_{19,1}=-S_{20}, $$ $$
 S_{18,2}=S_{20}+425 S_{8,8,4}-1904S_{7,7,6}-17S_{16,2,2},$$ $$
 S_{17,3}=\frac{51S_{16,2,2}-1275S_{8,8,4}+5712S_{7,7,6}-2S_{20}}{2},$$
 $$
 S_{16,4}=\frac{2S_{20}-64S_{16,2,2}+1700S_{8,8,4}-7616S_{7,7,6}}{2},$$
 $$
  S_{15,5}=\frac{75S_{16,2,2}-2125S_{8,8,4}-2S_{20}+9520S_{7,7,6}}{2},$$
  $$
 S_{14,6}=\frac{2S_{20}+2506S_{8,8,4}-11204S_{7,7,6}-84S_{16,2,2}}{2},
 $$ $$
 S_{13,7}=\frac{91S_{16,2,2}+12558S_{7,7,6}-2821S_{8,8,4}-2S_{20}}{2},
 $$ $$
 S_{12,8}=\frac{2S_{20}-96S_{16,2,2}+3056S_{8,8,4}-13536S_{7,7,6}}{2},
$$ $$
 S_{11,9}=\frac{99S_{16,2,2}-2S_{20}-3201S_{8,8,4}+14124S_{7,7,6}}{2},
 $$ $$
 S_{10,10}=\frac{2S_{20}-100S_{16,2,2}+3250S_{8,8,4}-14320S_{7,7,6}}{2}.
 $$
6)  Solution of Constraints of degree 22:

The invariant polynomials $S_{22}$, $S_{18,2,2}$, $S_{9,9,4}$ and
$S_{8,8,6}$ are considered to be independent  and the other
dependent invariant polynomials can be written in terms of them as
follows:
 $$ S_{8,7,7}=0,\hspace{3 mm}
 S_{9,7,6}=-2S_{8,8,6}, $$ $$
 S_{9,8,5}=-S_{8,8,6},\hspace{3 mm}
 S_{10,6,6}=2S_{8,8,6}, $$ $$
 S_{10,7,5}=3S_{8,8,6},\hspace{3 mm}
 S_{10,9,3}=-S_{9,9,4}, $$ $$
 S_{10,8,4}=S_{8,8,6}-2S_{9,9,4},\hspace{3 mm}
 S_{11,6,5}=-7S_{8,8,6}, $$ $$
 S_{11,7,4}=2S_{9,9,4}-4S_{8,8,6},$$ $$
 S_{11,8,3}=3S_{9,9,4}-S_{8,8,6},\hspace{3 mm}
 S_{12,5,5}=7S_{8,8,6}, $$ $$
 S_{12,6,4}=11S_{8,8,6}+2S_{9,9,4},\hspace{3 mm}
 S_{12,7,3}=5S_{8,8,6}-5S_{9,9,4},$$ $$
 S_{13,5,4}=2S_{9,9,4}-25S_{8,8,6},\hspace{3 mm}
 S_{13,6,3}=7S_{9,9,4}-16S_{8,8,6},$$ $$
 S_{14,4,4}=25S_{8,8,6}-2S_{9,9,4},\hspace{3 mm}
 S_{14,5,3}=41S_{8,8,6}-9S_{9,9,4}, $$ $$
 S_{15,4,3}=13S_{9,9,4}-91S_{8,8,6},\hspace{3 mm}
 S_{16,3,3}=91S_{8,8,6}-13S_{9,9,4},$$ $$
S_{17,3,2}=-S_{18,2,2},$$ $$
S_{16,4,2}=26S_{9,9,4}-182S_{8,8,6}+S_{18,2,2},$$ $$
 S_{15,5,2}=273S_{8,8,6}-39S_{9,9,4}-S_{18,2,2},$$ $$
 S_{14,6,2}=48S_{9,9,4}-314S_{8,8,6}+S_{18,2,2},$$ $$
 S_{13,7,2}=330S_{8,8,6}-55S_{9,9,4}-S_{18,2,2},$$ $$
 S_{12,8,2}=60S_{9,9,4}+S_{18,2,2}-335S_{8,8,6},$$ $$
 S_{11,9,2}=336S_{8,8,6}-63S_{9,9,4}-S_{18,2,2},$$ $$
 S_{10,10,2}=\frac{S_{18,2,2}+64S_{9,9,4}-336S_{8,8,6}}{2},$$ $$
 S_{11,10,1}=\frac{-64S_{9,9,4}+336S_{8,8,6}-S_{18,2,2}}{2}, $$
 $$
 S_{12,9,1}=\frac{190S_{9,9,4}+3S_{18,2,2}-1008S_{8,8,6}}{2}, $$
 $$
 S_{13,8,1}=\frac{-310S_{9,9,4}-5S_{18,2,2}+1678S_{8,8,6}}{2}, $$
 $$
 S_{14,7,1}=\frac{-2338S_{8,8,6}+7S_{18,2,2}+420S_{9,9,4}}{2}, $$
 $$
 S_{15,6,1}=\frac{2966S_{8,8,6}-9S_{18,2,2}-516S_{9,9,4}}{2}, $$
 $$
 S_{16,5,1}=\frac{11S_{18,2,2}+594S_{9,9,4}-3512S_{8,8,6}}{2}, $$
 $$
 S_{17,4,1}=\frac{3876S_{8,8,6}-13S_{18,2,2}-646S_{9,9,4}}{2}, $$
 $$
 S_{18,3,1}=\frac{-3876S_{8,8,6}+646S_{9,9,4}+15S_{18,2,2}}{2},
 $$ $$
 S_{19,2,1}=\frac{-646S_{9,9,4}-19S_{18,2,2}+3876S_{8,8,6}}{2},
 $$ $$
 S_{20,1,1}=\frac{-3876S_{8,8,6}+19S_{18,2,2}+646S_{9,9,4}}{2},
 $$ $$
 S_{21,1}=-S_{22}, $$ $$
 S_{20,2}=S_{22}-19S_{18,2,2}-646S_{9,9,4}+3876S_{8,8,6},$$ $$
 S_{19,3}=\frac{57S_{18,2,2}-2S_{22}+1938S_{9,9,4}-11628S_{8,8,6}}{2},
 $$ $$
 S_{18,4}=\frac{2S_{22}-72S_{18,2,2}-2584S_{9,9,4}+15504S_{8,8,6}}{2},
 $$ $$
 S_{17,5}=\frac{3230S_{9,9,4}+85S_{18,2,2}-19380S_{8,8,6}-2S_{22}}{2},
 $$ $$
 S_{16,6}=\frac{2S_{22}-96S_{18,2,2}-3824S_{9,9,4}+22892S_{8,8,6}}{2},
 $$ $$
 S_{15,7}=\frac{105S_{18,2,2}+4340S_{9,9,4}-25858S_{8,8,6}-2S_{22}}{2},
 $$ $$
S_{14,8}=\frac{28196S_{8,8,6}-112S_{18,2,2}+2S_{22}-4760S_{9,9,4}}{2},
$$ $$
 S_{13,9}=\frac{5070S_{9,9,4}+117S_{18,2,2}-29874S_{8,8,6}-2S_{22}}{2},
 $$ $$
 S_{12,10}=\frac{30882S_{8,8,6}+2S_{22}-120S_{18,2,2}-5260S_{9,9,4}}{2},
 $$ $$
 S_{11,11}=\frac{121S_{18,2,2}+4678S_{9,9,4}-2S_{22}-31218S_{8,8,6}}{4}.
  $$
{\large \appendix{ APPENDIX IV:the recursion relation and
$\lambda_{max}$ for $q\geq 6$}}
 a){\bf b=2}
{\bf q=6}
 $$ \left(\begin{array}{c}A_{6(n+1)} \\
A_{4,1,1(n+1)}\end{array}\right)=
\left(\begin{array}{cc}1.090534979 &   -.1646090535\\.1481481481 &
.1851851852 \end{array}\right) \left(\begin{array}{c}A_{6(n)} \\
A_{4,1,1(n)}\end{array}\right),$$\\ $$\lambda_{max}=1.062745991.
\hspace{5cm}  $$ {\bf q=8}
 $$ \left(\begin{array}{c}A_{8(n+1)} \\
A_{3,3,2(n+1)}\end{array}\right)= \left(\begin{array}{cc}
1.039323274  & .5121170553\\-.03840877915  &
.1742112483\end{array}\right) \left(\begin{array}{c}A_{8(n)} \\
A_{3,3,2(n)}\end{array}\right),$$ \\  $$
\lambda_{max}=1.015955376. \hspace{5cm}  $$ {\bf q=10}
 $$ \left(\begin{array}{c}A_{10(n+1)}
\\ A_{4,4,2(n+1)}\end{array}\right)= \left(\begin{array}{cc}
1.017375400  &   -.3840877915\\.02987349489  &  .1486053955
\end{array}\right) \left(\begin{array}{c}A_{10(n)} \\
A_{4,4,2(n)}\end{array}\right),$$ \\  $$
\lambda_{max}=1.003961045. \hspace{5cm}  $$
 {\bf q=12}
  $$ \left(\begin{array}{c}A_{12(n+1)}
\\ A_{8,2,2(n+1)}\\ A_{6,3,3(n+1)}\end{array}\right)=
\left(\begin{array}{ccc}1.007711110  &   -.1266744568  &
.1217068310 \\.05378583888  &   .1908753747   &  -.1417805551
\\.007315957933  &  .07681755830  &
-.04709647920\end{array}\right) \left(\begin{array}{c}A_{12(n)} \\
A_{8,2,2(n)}\\ A_{6,3,3(n+1)}\end{array}\right),$$ \\  $$
\lambda_{max}= 1.000983621.   \hspace{5cm}$$
 {\bf q=14}
$$ \left(\begin{array}{c}A_{14(n+1)} \\ A_{6,6,2(n+1)}\\
A_{5,5,4(n+1)}\end{array}\right)=\left(\begin{array}{ccc}
1.003425906   &   -.1557074696  &   .6579896295\\ .01048470103  &
-.09541040304  &  .5668362057\\-.001896729835  &  -.04335382479  &
.2332808346\end{array}\right) \left(\begin{array}{c}A_{14(n)} \\
A_{6,6,2(n)}\\ A_{5,5,4(n+1)}\end{array}\right),$$ \\  $$
\lambda_{max}= 1.000244974.   \hspace{5cm} $$
 {\bf q=16}
  $$
\left(\begin{array}{c}A_{16(n+1)} \\ A_{7,7,2(n+1)}\\
A_{6,6,4(n+1)}\end{array}\right)=\left(\begin{array}{ccc}
1.001522485  &    .09132402907   &  -.5053262942\\-.006679533152 &
-.1334905160  &   1.056414123 \\.001475234316   & -.03567658498  &
.2661499583 \end{array}\right) \left(\begin{array}{c}A_{16(n)} \\
A_{7,7,2(n)}\\ A_{6,6,4(n+1)}\end{array}\right),$$ \\  $$
\lambda_{max}= 1.000061131. \hspace{5cm}$$
 {\bf q=18}
  $${\scriptsize
\left(\begin{array}{c}A_{18(n+1)} \\ A_{14,2,2(n+1)}\\
A_{7,7,4(n+1)}\\
A_{8,5,5(n+1)}\end{array}\right)=\left(\begin{array}{cccc}1.000676645
& -.02587988061 & -.5694995445 & .5946011699\\.04976341868 &
.3702685508 & 6.165437409 & -6.474053436\\-.0007861225946 &
-.004847198466 & -.1026962541 &.1087868226\\.0003612818732 &
.009483649173 & .1320936849 & -.1378233896\\1.000015270 &
.1122123554 & .01717707276 & .001020854222 \end{array}\right)
\left(\begin{array}{c}A_{18(n)} \\ A_{14,2,2(n)}\\
A_{7,7,4(n)}\\A_{8,5,5(n)}\end{array}\right)},$$ \\    $$
\lambda_{max}= 1.000015270.      \hspace{5cm} $$
 {\bf q=20}
 $$ {\scriptsize
\left(\begin{array}{c}A_{20(n+1)} \\ A_{16,2,2(n+1)}\\
A_{8,8,4(n+1)}\\
A_{7,7,6(n+1)}\end{array}\right)=\left(\begin{array}{cccc}1.000300729
& -.01428439338 & .4535545013 &  -2.007507524\\.04955188625 &
.4249966010 & -10.27981253 & 46.05861448\\.0005177630138 &
.0009212316079 & -.05329922806 & .2455943625\\-.00009366567084 &
-.002585841556 & .05473587640 & -.2437584270
\end{array}\right) \left(\begin{array}{c}A_{20(n)} \\
A_{16,2,2(n)}\\ A_{8,8,4(n)}\\A_{7,7,6(n)}\end{array}\right)},$$
\\  $$ \lambda_{max}= 1.000003815.     \hspace{5cm} $$
 {\bf q=22}
  $$ {\scriptsize
\left(\begin{array}{c}A_{22(n+1)} \\ A_{18,2,2(n+1)}\\
A_{9,9,4(n+1)}\\
A_{8,8,6(n+1)}\end{array}\right)=\left(\begin{array}{cccc}1.000133657
& -.007718672824 & -.3331935728 & 1.963553834\\.04945789936 &
.4775535736 & 15.78748967 & -94.73506749\\-.0003298534890 &
.001784479705 & .02202297308 & -.1217001716\\.00007285107732 &
.002090751581 & .06241497465 & -.3728127167\end{array}\right)
\left(\begin{array}{c}A_{22(n)} \\ A_{18,2,2(n)}\\
A_{9,9,4(n)}\\A_{8,8,6(n)}\end{array}\right)},$$ \\  $$
\lambda_{max}= 1.000000954.    \hspace{5cm}$$
 b){\bf b=3}
{\bf q=6} $$ \left(\begin{array}{c}A_{6(n+1)} \\
A_{4,1,1(n+1)}\end{array}\right)=
\left(\begin{array}{cc}1.082584637 &   -.07343878826\\.1808430161
&   .04755671531 \end{array}\right) \left(\begin{array}{c}A_{6(n)}
\\ A_{4,1,1(n)}\end{array}\right),$$
\\ $$\lambda_{max}= 1.069590059.   \hspace{5cm}$$
{\bf q=8}
 $$ \left(\begin{array}{c}A_{8(n+1)} \\
A_{3,3,2(n+1)}\end{array}\right)= \left(\begin{array}{cc}
1.025056713  &   .1818484281 \\-.04533096632  &
.03647844913\end{array}\right) \left(\begin{array}{c}A_{8(n)} \\
A_{3,3,2(n)}\end{array}\right),$$ \\  $$
\lambda_{max}=1.016646559. \hspace{5cm} $$
 {\bf q=10}
 $$ \left(\begin{array}{c}A_{10(n+1)} \\
A_{4,4,2(n+1)}\end{array}\right)=
\left(\begin{array}{cc}1.007845912 &   -.1085523780\\.03420274178
&  .02815706873\end{array}\right) \left(\begin{array}{c}A_{10(n)}
\\ A_{4,4,2(n)}\end{array}\right),$$
\\  $$ \lambda_{max}= 1.004041374.   \hspace{5cm}$$
{\bf q=12}
 $$ \left(\begin{array}{c}A_{12(n+1)} \\ A_{8,2,2(n+1)} \\
A_{6,3,3(n+1)}\end{array}\right)=
\left(\begin{array}{ccc}1.002501314  &   -.02774717658 &
.02388825224 \\.06022405661  &  .05660671065   &
-.04390015579\\.01079114607  &  .01664662055  &   -.01252511118
\end{array}\right) \left(\begin{array}{c}A_{12(n)} \\ A_{8,2,2(n)}\\
A_{6,3,3(n+1)}\end{array}\right),$$ \\  $$  \lambda_{max}=
1.001026135. \hspace{5cm} $$
 {\bf q=14}
  $$
\left(\begin{array}{c}A_{14(n+1)} \\ A_{6,6,2(n+1)} \\
A_{5,5,4(n+1)}\end{array}\right)=\left(\begin{array}{ccc}
1.000805713  &    -.02587192638   &  .1198110012 \\.008633787717 &
-.02491545222   &  .1279091010 \\-.002716116832  & -.009331353906
&  .04723947173\end{array}\right) \left(\begin{array}{c}A_{14(n)}
\\ A_{6,6,2(n)}\\ A_{5,5,4(n+1)}\end{array}\right),$$ \\  $$ \lambda_{max}=
1.000246359. \hspace{5cm} $$
 {\bf q=16}
  $$
\left(\begin{array}{c}A_{16(n+1)} \\ A_{7,7,2(n+1)} \\
A_{6,6,4(n+1)}\end{array}\right)=\left(\begin{array}{ccc}
1.000261069    &   .01135511506  &   -.07104143816\\-.004468408282
&  -.03190697902   &   .2276301973 \\.002050937173  &
-.007546589748  &  .05339121346  \end{array}\right)
\left(\begin{array}{c}A_{16(n)} \\ A_{7,7,2(n)}\\
A_{6,6,4(n+1)}\end{array}\right),$$ \\  $$ \lambda_{max}=
1.000061317. \hspace{5cm} $$
 {\bf q=18}
  $$ {\scriptsize
\left(\begin{array}{c}A_{18(n+1)} \\ A_{14,2,2(n+1)}\\
A_{7,7,4(n+1)}\\
A_{8,5,5(n+1)}\end{array}\right)=\left(\begin{array}{cccc}1.000084877
& -.002388883753 & -.04740112573 & .04969629320
\\.06005002155 & .07399773075 & 1.274170117 &
-1.337882220\\-.0009014749149 & -.0004616094519 & -.008788472030 &
.009233751643 \\ .0006471907605 & .002494396769 & .04199894504 &
-.04409378714 \end{array}\right) \left(\begin{array}{c}A_{18(n)}
\\ A_{1422(n)}\\ A_{7,7,4(n)}\\A_{8,5,5(n)}\end{array}\right)},$$
\\  $$ \lambda_{max}= 1.000015297.      \hspace{5cm} $$ {\bf q=20}
 $$  {\scriptsize \left(\begin{array}{c}A_{20(n+1)} \\ A_{16,2,2(n+1)}\\
A_{8,8,4(n+1)}\\
A_{7,7,6(n+1)}\end{array}\right)=\left(\begin{array}{cccc}1.000027647
& -.0009744218073 & .02806977909 & -.1253211808\\.05999917318 &
.08477443355 & -2.107893057 & 9.443400292\\.0005178070963 &
-.0004282216637 & .009448647802 & -.04228755067\\-.0001628983371 &
-.0006696634259 & .01635654929 & -.07326845888
\end{array}\right) \left(\begin{array}{c}A_{20(n)}
\\ A_{16,2,2(n)}\\
A_{8,8,4(n)}\\A_{7,7,6(n)}\end{array}\right)},$$ \\ $$
\lambda_{max}= 1.000003819.     \hspace{5cm}$$
 {\bf q=22}
  $$ {\scriptsize
\left(\begin{array}{c}A_{22(n+1)} \\ A_{18,2,2(n+1)}\\
A_{9,9,4(n+1)}\\
A_{8,8,6(n+1)}\end{array}\right)=\left(\begin{array}{cccc}1.000006761
& -.0002279253840 & -.009315026611 & .05551458536\\.05998283956 &
.09512343412 & 3.219304290 & -19.31589634\\-.0002679923170 &
.0009873502040 & .03199666409 & -.1919144288\\ .0001230048643 &
.0005306869789 & .01770755556 & -.1062348858\end{array}\right)
\left(\begin{array}{c}A_{22(n)}
\\ A_{18,2,2(n)}\\
A_{9,9,4(n)}\\A_{8,8,6(n)}\end{array}\right)},$$
\\ $$ \lambda_{max}=  1.000002379.    \hspace{5cm} $$
 c){\bf b=4}
{\bf q=6} $$ \left(\begin{array}{c}A_{6(n+1)} \\
A_{4,1,1(n+1)}\end{array}\right)= \left(\begin{array}{cc}
1.076310660  &  -.03365228885\\.1889456076 &   .01616947736
\end{array}\right) \left(\begin{array}{c}A_{6(n)} \\
A_{4,1,1(n)}\end{array}\right),$$ \\ $$\lambda_{max}=1.070278597.
\hspace{5cm} $$ {\bf q=8}
 $$ \left(\begin{array}{c}A_{8(n+1)} \\
A_{3,3,2(n+1)}\end{array}\right)= \left(\begin{array}{cc}
1.019524120  &   .06535850011\\-.05434716964 &   371.8425102
\end{array}\right) \left(\begin{array}{c}A_{8(n)} \\
A_{3,3,2(n)}\end{array}\right),$$ \\  $$ \lambda_{max}=1.0195338.
\hspace{5cm} $$
\\ \\
 {\bf q=10}
  $$ \left(\begin{array}{c}A_{10(n+1)} \\
A_{4,4,2(n+1)}\end{array}\right)=
\left(\begin{array}{cc}1.005151073 &   -.03282144763\\.03495279823
& .007384841623\end{array}\right) \left(\begin{array}{c}A_{10(n)}
\\ A_{4,4,2(n)}\end{array}\right),$$
\\  $$ \lambda_{max}= 1.003999975. \hspace{5cm}$$
{\bf q=12}
 $$ \left(\begin{array}{c}A_{12(n+1)} \\ A_{8,2,2(n+1)} \\
A_{6,3,3(n+1)}\end{array}\right)=
\left(\begin{array}{ccc}1.001388320  &   -.007332723661  &
.006026327075 \\.06394351082  &  .009657753638   &
-.007508055989\\ .01147176558  &  .004507477340  &
-.003470225654\end{array}\right) \left(\begin{array}{c}A_{12(n)}
\\ A_{8,2,2(n)}\\ A_{6,3,3(n+1)}\end{array}\right),$$ \\  $$ \lambda_{max}=
1.000986557.   \hspace{5cm} $$
\\ \\
 {\bf q=14}
  $$
\left(\begin{array}{c}A_{14(n+1)} \\ A_{6,6,2(n+1)} \\
A_{5,5,4(n+1)}\end{array}\right)=\left(\begin{array}{ccc}
1.000380234   &   -.006079338806  &  .02938006638\\.008215782561 &
-.006916822481  &  .03482160520\\-.002872694778  & -.002513285925
&  .01260695484\end{array}\right) \left(\begin{array}{c}A_{14(n)}
\\ A_{6,6,2(n)}\\ A_{5,5,4(n+1)}\end{array}\right),$$ \\  $$ \lambda_{max}=
1.000245203. \hspace{5cm} $$
 {\bf q=16}
  $$
\left(\begin{array}{c}A_{16(n+1)} \\ A_{7,7,2(n+1)} \\
A_{6,6,4(n+1)}\end{array}\right)=\left(\begin{array}{ccc}
1.000105428   &   .002387138955   &  -.01597611501\\-.004013610557
&  -.008712843003  &  .06128427304 \\.002158295483  &
-.002024493055  &  .01421013434 \end{array}\right)
\left(\begin{array}{c}A_{16(n)} \\ A_{7,7,2(n)}\\
A_{6,6,4(n+1)}\end{array}\right),$$ \\  $$ \lambda_{max}=
1.000061148. \hspace{5cm} $$ {\bf q=18}
 $$ {\scriptsize \left(\begin{array}{c}A_{18(n+1)} \\ A_{14,2,2(n+1)}\\
A_{7,7,4(n+1)}\\
A_{8,5,5(n+1)}\end{array}\right)=\left(\begin{array}{cccc}1.000029509
& -.0004497502350 & -.008243519261 & .008652383249\\.06187422841
&.01967111602 & .3403871980 & -.3574066079\\-.0009123885980 &
-.00009719487958 & -.001741086261 &.001828250148\\.0007092075724 &
.0006921244620 & .01190983853 & -.01250523720 \end{array}\right)
\left(\begin{array}{c}A_{18(n)}
\\ A_{14,2,2(n)}\\
A_{7,7,4(n)}\\A_{8,5,5(n)}\end{array}\right)},$$ \\  $$
\lambda_{max}= 1.000015271.      \hspace{5cm} $$
 {\bf q=20}
 $$ {\scriptsize
\left(\begin{array}{c}A_{20(n+1)} \\ A_{16,2,2(n+1)}\\
A_{8,8,4(n+1)}\\
A_{7,7,6(n+1)}\end{array}\right)=\left(\begin{array}{cccc}1.000008319
& -.0001645061367 & .004348107103 & -.01946561802\\.06184409730 &
.02252802695 & -.5624149410 & 2.519619600\\.0005078921879 &
-.0001439100811 & .003508323524 & -.01571652660\\-.0001776181030
& -.0001853563027 & .004606804705 & -.02063832629
\end{array}\right) \left(\begin{array}{c}A_{20(n)}
\\ A_{16,2,2(n)}\\
A_{8,8,4(n)}\\A_{7,7,6(n)}\end{array}\right)},$$ \\  $$
\lambda_{max}= 1.000003788.     \hspace{5cm} $$
 {\bf q=22}
 $$ {\scriptsize
\left(\begin{array}{c}A_{22(n+1)} \\ A_{18,2,2(n+1)}\\
A_{9,9,4(n+1)}\\
A_{8,8,6(n+1)}\end{array}\right)=\left(\begin{array}{cccc}1.000002358
& -.00005834048617 & -.002108427732 & .01263785041\\.06183577589 &
.02527262136 & .8582479458 & -5.149488875\\-.0002481471736 &
.0002926869789 & .009839879848 & -.05903810198\\.0001334512767 &
.0001464026161 & .004954070521 & -.02972423554\end{array}\right)
\left(\begin{array}{c}A_{22(n)}
\\ A_{18,2,2(n)}\\
A_{9,9,4(n)}\\A_{8,8,6(n)}\end{array}\right)},$$ \\  $$
\lambda_{max}= 1.000000954.    \hspace{5cm} $$
 d){\bf b=5}
{\bf q=6}
 $$ \left(\begin{array}{c}A_{6(n+1)} \\
A_{4,1,1(n+1)}\end{array}\right)=
\left(\begin{array}{cc}1.044360741 &  -.01006023599\\.3156061304 &
.003157528970 \end{array}\right) \left(\begin{array}{c}A_{6(n)}
\\ A_{4,1,1(n)}\end{array}\right),$$
\\  $$ \lambda_{max}=1.041302331.   \hspace{5cm} $$
{\bf q=8}
 $$ \left(\begin{array}{c}A_{8(n+1)} \\
A_{3,3,2(n+1)}\end{array}\right)= \left(\begin{array}{cc}
1.009037038   &   .01421998195 \\.07691145227  &
-.005139564359\end{array}\right) \left(\begin{array}{c}A_{8(n)} \\
A_{3,3,2(n)}\end{array}\right),$$ \\  $$
\lambda_{max}=1.010114286. \hspace{5cm} $$ {\bf q=10}
 $$ \left(\begin{array}{c}A_{10(n+1)} \\
A_{4,4,2(n+1)}\end{array}\right)=
\left(\begin{array}{cc}1.001897065 & -.005511338227\\.03479355808
&  .002253185745
\end{array}\right) \left(\begin{array}{c}A_{10(n)} \\
A_{4,4,2(n)}\end{array}\right),$$ \\  $$
\lambda_{max}=1.001705201. \hspace{5cm} $$ {\bf q=12} $$
\left(\begin{array}{c}A_{12(n+1)} \\ A_{8,2,2(n+1)} \\
A_{6,3,3(n+1)}\end{array}\right)=
\left(\begin{array}{ccc}1.000405841  &   -.0009777920845  &
.0007888947248\\.06389106764  &   .002988237303  &
-.002324086713\\.01925523989   &  .001402362796   &
-.001089147707\end{array}\right) \left(\begin{array}{c}A_{12(n)}
\\ A_{8,2,2(n)}\\ A_{6,3,3(n+1)}\end{array}\right),$$ \\  $$ \lambda_{max}=
1.000358489.   \hspace{5cm}$$ {\bf q=14} $$
\left(\begin{array}{c}A_{14(n+1)} \\ A_{6,6,2(n+1)} \\
A_{5,5,4(n+1)}\end{array}\right)=\left(\begin{array}{ccc}
1.000087998  &   -.0006533768126  &  .003203625771 \\.008004412267
&  -.002154118594   &   .01079323997 \\.004778768070  &
.001472733806   &  -.007372115949\end{array}\right)
\left(\begin{array}{c}A_{14(n)} \\ A_{6,6,2(n)}\\
A_{5,5,4(n+1)}\end{array}\right),$$ \\ $$ \lambda_{max}=
1.000097981. \hspace{5cm} $$ {\bf q=16} $$
\left(\begin{array}{c}A_{16(n+1)} \\ A_{7,7,2(n+1)} \\
A_{6,6,4(n+1)}\end{array}\right)=\left(\begin{array}{ccc}
1.000019275   &  .0002082580899   &  -.001424787973\\.003961084604
&   .002614693987   &  -.01832747999 \\.002150223007  &
-.0006222811611  &  .004359625253 \end{array}\right)
\left(\begin{array}{c}A_{16(n)} \\ A_{7,7,2(n)}\\
A_{6,6,4(n+1)}\end{array}\right),$$ \\  $$ \lambda_{max}=
1.000017019. \hspace{5cm}$$ {\bf q=18} $$ {\scriptsize
\left(\begin{array}{c}A_{18(n+1)} \\ A_{1422(n+1)}\\
A_{774(n+1)}\\
A_{855(n+1)}\end{array}\right)=\left(\begin{array}{cccc}
1.000004256 & -.00003193265577 & -.0005685434366 &
.0005968699293\\.06193405435 & .006096146153 & .1056112005 &
-.1108917618\\.001017169377 & .00006930842110 & .001204750315 &
-.001264989081\\.001191712555 & .0002627970450 & .004548661710 &
-.004776093729 \end{array}\right) \left(\begin{array}{c}A_{18(n)}
\\ A_{14,2,2(n)}\\ A_{7,7,4(n)}\\A_{855(n)}\end{array}\right)},$$
\\  $$ \lambda_{max}= 1.000002408.     \hspace{5cm}$$ {\bf q=20}
 $$ {\scriptsize \left(\begin{array}{c}A_{20(n+1)} \\ A_{16,2,2(n+1)}\\
A_{8,8,4(n+1)}\\
A_{7,7,6(n+1)}\end{array}\right)=\left(\begin{array}{cccc}1.000000946
& -.000009482285099 & .0002434259750 & -.001090262082\\.06190933065
& .006979673785 & -.1744166676 & .7813866910\\.0004953876882 &
-.00004636993771 & .001150699865 & -.005155112958\\.0002958221549
& .0001041331110 & -.002597984312 & .01163895822
\end{array}\right) \left(\begin{array}{c}A_{20(n)}
\\ A_{16,2,2(n)}\\
A_{8,8,4(n)}\\A_{7,7,6(n)}\end{array}\right)},$$ \\  $$
\lambda_{max}= 1.000000141.    \hspace{5cm}$$ {\bf q=22} $$
{\scriptsize \left(\begin{array}{c}A_{22(n+1)} \\
A_{18,2,2(n+1)}\\ A_{9,9,4(n+1)}\\
A_{8,8,6(n+1)}\end{array}\right)=\left(\begin{array}{cccc}1.000000211
& -.000002740626906 & -.00009573710813 &
.0005742469550\\.06190294246 & .007829018982 & .2660888394 &
-1.596533072\\.0002452104252 & -.00008750002217 & .2660888394 &
.01778872116\\.0001330977107 & .00004502739760 & .001528690467 &
-.009172137322\end{array}\right) \left(\begin{array}{c}A_{22(n)}
\\ A_{18,2,2(n)}\\ A_{9,9,4(n)}\\A_{8,8,6(n)}\end{array}\right)},$$
$$ \lambda_{max}= 1.000000086.    \hspace{5cm}$$

\vspace{10mm} {\bf {\large  ACKNOWLEDGEMENT}}

We wish to thank  Dr. S. K. A. Seyed Yagoobi for his careful
reading the article and for his constructive comments.


\begin{thebibliography}{99}

\bibitem{Mandel}{\it B. B. Mandelbrot,} {\em The Fractal Geometry of Nature (Freeman, New York , 1982)}

\bibitem{Sch}{\it M. Schroeder,} {\em Fractals, Chaos, Power laws, Minutes from an Infinite Paradise (Freeman, New York, 1991).}

\bibitem{Gefen}{\it Y. Gefen, A. Aharony, B. B. Mandelbrot and S. Kirkpatrich}; {\em Physics Review Letter {\bf 47} (1981) 1771.}

\bibitem{Cle}{\it J. P. Clerc, G. Giraud, J. M. Laugier and J.M.Luck} {\em  Advances in Physics, vol {\bf 39} no 3 (1990) 191.}

\bibitem{Ste}{\it Stephane Roux and Catalin D. Mitescu,} {\em Physical review {\bf B} {\bf 35} (1987) 898.}

\bibitem{Pak}{\it Pak-Yee Tong and Kin-Wah Yu,} {\em  Physics letters {\bf A} {\bf 160} (1991) 293.}

\bibitem{Jul}{\it R. Julllian and R. Botet,} {\em Aggregation and Fractal Aggregates (World Scientific, Singapore, 1987).}

\bibitem{Jaf}{\it M.A. Jafarizadeh and S.K.A. Seyed-Yagoobi, } {\em Physics Letters {\bf A}{\bf 192} (1994) 109.}

\bibitem{Jafa}{\it M.A. Jafarizadeh and S.K.A. Seyed-Yagoobi, } {\em  J. Polymer Science Part B: Polymer Physics 34, John Wiley $\&$ Sons Inc. (1996)}

\bibitem{Jafar}{\it M.A. Jafarizadeh and S.K.A. Seyed-Yagoobi, } {\em Indian Journal of Physics {\bf70} B 2 (1996) 751.}

\bibitem{Jafar1}{\it M.A. Jafarizadeh and S.K.A. Seyed-Yagoobi, } {\em Europian Physical Journal B 7(1999) 429.}

\bibitem{Jafar2}{\it M.A. Jafarizadeh, } {\em Europian Physical Journal B 4(1998) 103.}

\bibitem{Jafar3}{\it M.A. Jafarizadeh, } {\em Restoration of macroscopic isotropy on $d+1$-simplex fractal
conductors, cond-mat/0005227(2000) to be published in Physica A .}

\bibitem{Andr}{\it George E. Andrews,} {\em The Theory of Partitions (Addison-Wesley, London , 1976).}
\newpage
{\bf Figures Captions}\\Figur 1. Sierpinsky fractal with decimation number b=3,
 prtitions of 2 denote the subfractals and partitions of 3 indicate the vertices
, respectively.\\ Figure 2. Sierpinski fractal resistor networks
with decimation number b=3.
\end{thebibliography}
\end{document}